\documentclass[useAMS,usenatbib]{mn2e}

\usepackage{amsmath}

\newcommand\bfa{{\bf a}}
\newcommand\bfu{{\bf u}}
\newcommand\bfb{{\bf b}}

\newcommand\bfz{{\bf z}}
\newcommand\bfB{{\bf B}}
\newcommand\bfG{{\bf G}}
\newcommand\bfX{{\bf X}}
\newcommand\calE{{\cal E}}
\newcommand\bfcalE{\boldsymbol{\cal E}}

\title[Problems with kinematic mean field electrodynamics]{Problems with kinematic mean field electrodynamics at high magnetic Reynolds numbers}
\author[F. Cattaneo and D.W. Hughes]{F. Cattaneo$^{1}$\thanks{E-mail:
cattaneo@flash.uchicago.edu; d.w.hughes@leeds.ac.uk} and D.W. Hughes$^{2}$\footnotemark[1]\\
$^{1}$Department of Astronomy and Astrophysics and the Computation Institute, University of Chicago, Chicago, IL 60637, USA\\
$^{2}$Department of Applied Mathematics, University of Leeds, Leeds LS2 9JT, UK}
\begin{document}

\date{\today}

\pagerange{\pageref{firstpage}--\pageref{lastpage}} \pubyear{2002}

\maketitle

\label{firstpage}

\begin{abstract}
We discuss the applicability of the kinematic $\alpha$-effect formalism at high magnetic Reynolds numbers. In this regime the underlying flow is likely to be a small-scale dynamo, leading to the exponential growth of fluctuations. Difficulties arise with both the actual calculation of the $\alpha$ coefficients and with its interpretation. We argue that although the former may be circumvented --- and we outline several procedures by which the the $\alpha$ coefficients can be computed in principle --- the interpretation of these quantities in terms of the evolution of the large-scale field may be fundamentally flawed.
\end{abstract}


\section{Introduction}

Mean field electrodynamics was formalised over forty years ago in a remarkable paper by Steenbeck, Krause \& R\"adler (1966). Within its framework it is possible to derive an equation for the evolution of a magnetic field on a scale large compared with that of the velocity. This equation is much simpler to solve than the induction equation from which it is derived. Consequently, it has had an enormous influence on dynamo theory to this day.

Strictly speaking, mean field electrodynamics is a kinematic theory that addresses the growth of a weak seed field. Its function is thus to predict the growth rate and structure of the generated magnetic field. In its simplest form, where the underlying velocity is isotropic and homogeneous, the evolution equation for the large-scale field depends on two quantities; $\alpha$, the mean induction, and $\beta$, the turbulent diffusivity. The growth rate of a field of length scale $1/k$ is then 
\begin{equation}
s = \alpha k - \beta k^2.
\label{eq:grate}
\end{equation}
(Moffatt 1978). The beauty of this result is that provided $\alpha$ is non-zero, dynamo growth is guaranteed on sufficiently large scales. In the kinematic limit the coefficients $\alpha$ and $\beta$ are determined solely by the properties of the velocity and by the magnetic Reynolds number $Rm$.

Mean field electrodynamics relies on a scale separation between fluctuating and mean quantities. Provided such a scale separation can be enforced, the formalism is valid for any value of $Rm$. One of the challenges of the theory is to calculate $\alpha$ and $\beta$ in terms of properties of the velocity. If $Rm$ is small this can be rigorously accomplished by use of what is known as the first order smoothing approximation (FOSA); if $Rm$ is large, the determination of the coefficients $\alpha$ and $\beta$ becomes more problematic. Indeed there has been considerable discussion over whether results obtained under FOSA can be extended to the high $Rm$ regime, where the approximation is not valid (see Moffatt 1978, Krause \& R\"adler 1980). However, we believe that there is a more fundamental problem associated with the high $Rm$ regime, namely the exponential growth of magnetic fluctuations (small-scale dynamo action). In this case scale separation cannot be maintained and the whole idea of mean field theory breaks down. In this paper we argue that even if the coefficients $\alpha$ and $\beta$ can be determined they fail to provide information about the rate of growth of the large-scale field.

In \S2 we give a brief outline of the formulation of mean field electrodynamics and the derivation of the $\alpha$ and $\beta$ coefficients. In \S3 we introduce the various methods that are used to determine $\alpha$ and $\beta$. In \S4 we discuss convergence and issues relating to the influence of initial conditions and the requirements on sample size in order to achieve a given accuracy. In \S5 we discuss the problems that arise with the physical interpretation of the 
$\alpha$ and $\beta$ coefficients at high $Rm$.

\section[]{Formulation of Mean Field Electrodynamics}

The motivation for mean field electrodynamics is to understand the evolution of magnetic fields on scales large compared with those of the turbulent velocity responsible for its generation. The starting point is the magnetic induction equation,
\begin{equation}
\frac{\partial \bfB}{\partial t} = \nabla \times (\bfu \times \bfB) + \eta \nabla^2 \bfB,
\label{eq:ind}
\end{equation}
where $\bfB$ is the magnetic field and $\bfu$ the velocity. The idea is to introduce an average over spatial scales intermediate between the large integral scales and the small scales characteristic of the velocity, so that 
\begin{equation}
\bfB = \langle \bfB \rangle + \bfb ;
\label{eq:meanfluc}
\end{equation}
we shall assume that the velocity has no large scale. Taking the average of (\ref{eq:ind}) leads to the induction equations for the mean and fluctuating components of $\bfB$, namely
\begin{equation}
\frac{\partial \langle \bfB \rangle}{\partial t} = \nabla \times \bfcalE  + \eta \nabla^2 \langle \bfB \rangle,
\label{eq:mean_ind}
\end{equation}
\begin{equation}
\frac{\partial \bfb}{\partial t} = \nabla \times (\bfu \times \langle \bfB \rangle ) + \nabla \times \bfG + \eta 
\nabla^2 \bfb,
\label{eq:fluc_ind}
\end{equation}
where $\bfcalE = \langle \bfu \times \bfB \rangle$ is the mean electromotive force (emf), and $\bfG = \bfu \times \bfb - \langle \bfu \times \bfb \rangle$. In order to make progress with equation (\ref{eq:mean_ind}) one needs to relate $\bfcalE$ to $\langle \bfB \rangle$. In the kinematic limit, in which $\bfu$ is independent of $\bfB$, this can be achieved by noting that the linearity of equation (\ref{eq:fluc_ind}) leads to a linear expression of the form (Moffatt 1978)
\begin{equation}
\calE_i = \alpha_{ij} \langle B \rangle_j + \beta_{ijk} \frac{\partial}{\partial x_k}\langle B \rangle_j + \ldots ,
\end{equation}
where $\alpha_{ij}$ and $\beta_{ijk}$ are pseudo-tensors dependent on the properties of the velocity $\bfu$ and on $\eta$. 

The simplest case to consider, which still captures the essence of the problem, is that in which the velocity is homogeneous and isotropic and hence $\alpha_{ij}$ and $\beta_{ijk}$ take the form $\alpha_{ij} = \alpha \delta_{ij}$ and $\beta_{ijk} = \beta \epsilon_{ijk}$. Here $\alpha$ is a pseudo-scalar and hence is non-zero only for turbulence lacking reflectional symmetry; $\beta$, on the other hand, is a true scalar and hence can be non-zero even for reflectionally symmetric turbulence. Substituting $\alpha$ and $\beta$ into equation 
({\ref{eq:mean_ind}) gives
\begin{equation}
\frac{\partial \langle \bfB \rangle}{\partial t} = \alpha \nabla \times \langle \bfB \rangle  + (\eta +\beta)
\nabla^2 \langle \bfB \rangle .
\label{eq:mean_ind2}
\end{equation}
The interpretation of $\alpha$ and $\beta$ is straightforward: $\alpha$ represents mean induction and $\beta$ a turbulent diffusivity. The simplest solution of this equation can be obtained for the case when $\nabla \times \langle \bfB \rangle = k \langle \bfB \rangle$, in which case the growth rate is given by equation (\ref{eq:grate}).

It is easy to show that if the magnetic Reynolds number $Rm$ is small then the $\bfG$ term in equation (\ref{eq:fluc_ind}) can be neglected. \footnote{It can also be neglected if the correlation time of the velocity is sufficiently small; here though we shall mainly be concerned with the case when it is comparable with the turnover time.} In this case the fluctuations in the magnetic field arise solely from interactions between the velocity and the mean magnetic field; thus for the case when the mean field $\langle \bfB \rangle$ is uniform (and hence constant in time) these are bounded by some power of $Rm$. Solutions of equation (\ref{eq:fluc_ind}) can be readily obtained, giving explicit expressions for $\alpha$ and $\beta$. The practice of neglecting the $\bfG$ term is often referred to as the quasi-linear approximation or the first order smoothing approximation.

When $Rm$ is large one is no longer justified in neglecting the $\bfG$ term. In this regime there are no closed form 
solutions for $\bfb$ and one typically has to rely on numerical solutions of (\ref{eq:fluc_ind}). Once the solutions for $\bfb$ are determined then the coefficients $\alpha$ and $\beta$ can be reconstructed. The problem becomes particularly acute when small-scale dynamo action occurs and the fluctuations in $\bfb$ grow exponentially. Later we shall argue that the onset of small-scale dynamo action does not just introduce technical difficulties into the determination of the coefficients $\alpha$ and $\beta$ but does in fact undermine the assumption of scale separation and hence the very foundation of the mean field approach.

\section[]{Techniques for Calculating $\alpha$ and $\beta$}

In order to fix ideas let us consider a turbulent isotropic homogeneous flow, which may or may not be helical, at high $Rm$ in a periodic domain. We assume further that the velocity has a well-defined correlation length $\ell$ and, where necessary, we consider domains of size much greater than $\ell$. For such a system it is important to ask what can be measured in terms of $\alpha$ and $\beta$ and, crucially, what can be inferred about the initial growth of a magnetic field on large scales. 

In this section we shall discuss in detail various approaches that have been used in the determination of $\alpha$ and $\beta$. The first, and most natural, is to consider an experiment in which a uniform magnetic field $\bfB_0$ is applied, in the $z$-direction, say. In this case
\begin{equation}
\alpha = \frac{\langle \bfu \times \bfB \rangle \cdot {\hat \bfz}}{B_0} ,
\label{eq:alpha_unif}
\end{equation}
where $\langle \cdot \rangle$ denotes a volume average. The necessary size of the volume will be discussed presently. We note that since $\bfu$ has zero average then $\bfB$ can be replaced by $\bfb$ in equation (\ref{eq:alpha_unif}). The attractive feature of this procedure is that since $\bfB_0$ is uniform then scale separation is guaranteed. Clearly though this method can be used only to determine $\alpha$ but not $\beta$.

In order also to determine $\beta$ this approach can be modified by considering large-scale fields with non-zero gradient. From the definition
\begin{equation}
\bfcalE =
\langle \bfu \times \bfb \rangle =
\alpha \langle \bfB \rangle
- \beta \langle \nabla \times \bfB \rangle
\label{eq:emfab}
\end{equation}
it is clear that $\alpha$ and $\beta$ can be determined through the consideration of two independent large-scale magnetic fields yielding two independent measures of $\bfcalE$. Here $\bfb$ and $\langle \bfB \rangle$ are defined according to expression (\ref{eq:meanfluc}) and $\bfB$ is obtained from solution of the full induction equation 
(\ref{eq:ind}). A variant of this method, sometimes referred to as the test field procedure, is obtained by taking $\langle \bfB \rangle$ to be a fixed (i.e.\ time-independent) test field and $\bfb$ to be the solution of the fluctuating equation (\ref{eq:fluc_ind}) driven by that specific test field (Schrinner \textit{et al.}\ 2005).

An altogether different procedure can be constructed by considering a Lagrangian approach. If $\bfX ( \bfa ,t)$ represents the position at time $t$ of a fluid element initially located at position $\bfa$ then 
\begin{equation}
{\dot X_i}(\bfa, t) = u_i(\bfX (\bfa ,t),t) ,
\label{eq:traj}
\end{equation}
where $\bfu$ is the (Eulerian) fluid velocity. The evolution of an infinitesimal line element along the trajectory is given by
\begin{equation}
\delta x_i(\bfa, t) = J_{ij}(\bfa, t) \delta x_j (\bfa,0),
\label{eq:deltax}
\end{equation}
where $J_{ij}$ is the Jacobian satisfying
\begin{equation}
{\dot J_{ij}} = \frac{\partial u_i}{\partial x_k} \left( \bfX(\bfa , t) \right)  J_{kj} , \quad
J_{ij}(\bfa,0) = \delta_{ij}.
\label{eq:Jac}
\end{equation}
The correspondence between expression (\ref{eq:deltax}) and the Cauchy solution for a magnetic field in a perfect conductor (infinite $Rm$) led Moffatt (1974) to derive expressions for $\alpha$ and $\beta$ in terms of Lagrangian quantities, formally valid when $Rm$ is infinite. The expression for $\alpha$ is given by
\begin{equation}
\alpha = \frac{1}{3} \epsilon_{ijk} \langle u_i(\bfX , t) J_{jk}(t) \rangle .
\label{eq:lag_alpha}
\end{equation}
The corresponding expression for $\beta$ is somewhat involved and can be found as equation (7.116) in Moffatt (1978). The average here is over trajectories, which can be realised in two ways. One is for fixed $\bfa$ and many realisations of the velocity; the other is for many different values of $\bfa$ for a single realisation of the velocity. The former can be regarded as an ensemble average, whereas the latter can be regarded as a volume average. This procedure can be extended to the case of finite $Rm$ by adding a randomly fluctuating delta-correlated velocity to equation (\ref{eq:traj}) (Drummond \& Horgan 1986).

\section[]{Influence of Initial Conditions and Sample Size}

All of the procedures above involve evolving the fluctuating magnetic field to a certain time and then taking either a volume average or an average over trajectories. It is important that the average is taken after sufficient time has elapsed such that there is no influence of the initial conditions. This will typically take a few turnover times, the precise value depending on the initial condition for $\bfb$, on the velocity $\bfu$ and on $Rm$. Two cases must be distinguished. In one case $Rm$ is below the threshold for small-scale dynamo action; here the volume integrals involved in the measurement of the emf must be taken only after $\langle \bfb^2 \rangle$, say, has become stationary. In the other, the small-scale dynamo is operative and eventually will cause the fluctuations to grow exponentially with a well defined growth rate $s$; here the measurement can be taken only after $\langle \bfb^2 \rangle \exp(-2st)$ has become stationary.

The above considerations apply equally to either the Eulerian or Lagrangian methods provided $Rm$ is finite. In the Lagrangian approach with no stochastic component ($Rm$ infinite) there are no eigenfunctions and the different moments of $\bfb$ will grow at different rates. A possible strategy is to integrate until the flux shows a well-defined exponential growth. Interestingly, according to the flux conjecture (Finn \& Ott 1990), this growth rate is identical to that of the fast dynamo growth rate in the limit as $Rm \rightarrow \infty$.

The size of the volume over which averages are taken is also crucially important since ultimately it determines the size of the error. In order to estimate the required volume we make use of the relation
\begin{equation}
N \sim \left( \frac{\sigma}{\epsilon} \right)^2 ,
\label{eq:N}
\end{equation}
where $\sigma$ is the standard deviation, $\epsilon$ is the desired accuracy, measured, say, by the size of the error bars, and $N$ is the number of uncorrelated contributions to the average. If the turbulent velocity has a characteristic eddy size $\ell$ then it is reasonable to assume that the emf has a comparable characteristic scale; 
$N$ then denotes the number of patches of size $\ell$ that are needed to achieve the required accuracy. To proceed further we need an estimate for $\sigma$. For the evaluation of $\alpha$ it is reasonable to assume that
\begin{equation}
\sigma_\alpha \sim \frac{| \bfu \times \bfb |}{| \langle \bfB \rangle |}
\sim | \bfu | \frac{| \bfb |}{| \langle \bfB \rangle |} ,
\label{eq:sigma_alpha}
\end{equation}
where $| \bfu \times \bfb |$ and $| \bfu |$ refer to typical values of these quantities. If there is no small-scale dynamo action then 
\begin{equation}
\frac{| \bfb |}{| \langle \bfB \rangle |} \sim Rm^\gamma_1 ,
\label{eq:gamma_1}
\end{equation}
where $1/2 < \gamma_1 < 1$, its value depending on the velocity. This can be readily converted into an estimate for the linear size of the domain needed for accurate computation of the averages, namely
\begin{equation}
L \sim \ell \left( \frac{|\bfu| Rm^{\gamma_1}}{\epsilon}\right)^{2/3} .
\label{eq:L1}
\end{equation}
The situation becomes more complicated when small-scale dynamo action occurs and $|\bfu \times \bfb |$ increases exponentially at the small-scale dynamo growth rate. For the cases where the large-scale field is either uniform or held constant (the test field case) the contributions to the average grow exponentially with the small-scale dynamo growth rate. Here the requisite linear size of the domain takes the form
\begin{equation}
L \sim \ell \left( \frac{|\bfu| Rm^{\gamma_2} e^{st}}{\epsilon}\right)^{2/3} ,
\label{eq:L2}
\end{equation}
for some $O(1)$ exponent $\gamma_2$.

In the case when $\langle \bfB \rangle$ is itself time-dependent the estimate for $L$ depends on the magnitude of the ratio of the fluctuating emf to that of the mean field, both of which are growing exponentially. With this method, both the fluctuating field $\bfb$ and the mean field $\langle \bfB \rangle$ are obtained from the solution of the induction equation (\ref{eq:ind}); there is a single dynamo growth rate and therefore the fluctuating and mean fields will eventually grow at exactly the same rate. In this case $L$ is similar to that in (\ref{eq:L1}) but with a different $O(1)$ exponent, $\gamma_3$.

The above discussion has been couched in terms of volume averages. For the cases when the contributions to the averages are bounded, one can conceive of combining volume and time averages. However, when the contributions are growing exponentially, such a procedure is not tenable.

For the Lagrangian method the equivalent issue concerns the number of independent trajectories needed to determine $\alpha$. The source of concern here is that in (\ref{eq:lag_alpha}) the Jacobian on average grows exponentially with a rate given by the largest Lyapunov exponent. So, once again, the number of independent trajectories needed in the averaging procedure grows exponentially.

\section[]{What does it all mean?}

The considerations above suggest that the coefficients $\alpha$ and $\beta$ can, at least in principle, always be computed. However, in practice, the size of the computational domain (or the number of trajectories) required may be extremely large, and indeed, in some cases when small-scale dynamo action is present, may even be increasing exponentially in time. 

With the exception of the test field model all methods (Eulerian and Lagrangian) are based on averages of the exact solution of the induction equation, and therefore should yield the same results. The test field procedure, by contrast, invokes a predetermined arbitrary mean field. The quantities $\alpha$ and $\beta$ calculated through this technique will therefore be approximations to the true values of $\alpha$ and $\beta$, whose quality will depend on how well the true mean field is approximated by the choice of test field.

We now turn to the question of how $\alpha$ and $\beta$ relate to the growth of a dynamo field. First consider the case when no small-scale dynamo action is observed. Here, $\alpha \ell / \beta$ (i.e.\ $k \ell$) must be small; for if it were not then, by equation~(\ref{eq:grate}), one would observe dynamo action on a scale $\ell$ --- contrary to our assumption. One can conceive of a sequence of experiments of varying spatial extent $L$. When $L \sim \ell$ then, by assumption, no dynamo action is observed. As $L$ is increased, dynamo action will first set in when $L = \beta / \alpha$. Increasing $L$ further will lead to an increase in the growth rate until its maximum is reached at $L = 2 \beta / \alpha$. Further increases in $L$ will not lead to any further increases in the growth rate. So in the absence of small-scale dynamo action, everything is fine, and $\alpha$ and $\beta$ determine the growth rate of the observed magnetic structures.

Consider now the case when small-scale dynamo action is possible. What would be the outcome of repeating the experiments described above? By assumption, dynamo action takes place even when $L \sim \ell$. Furthermore, the dynamo growth rate will be independent of domain size. Finally, any average of the magnetic field on intermediate scales will grow at exactly the same rate. Crucially, the growth rate of the observed field has nothing to do with that predicted by equation~(\ref{eq:grate}). A particularly striking example can be seen for the case of a non-helical dynamo. Here $\alpha$ is zero and equation~(\ref{eq:grate}) therefore predicts that large-scale averages should decay exponentially at a scale-dependent rate, whereas, as we have just argued, any average will grow exponentially at the small-scale dynamo growth rate.

As argued above, the validity of the mean field approach breaks down when small scale dynamo action occurs, which one anticipates at high $Rm$. Interestingly, as pointed out by Moffatt (1978), there are problems in the high $Rm$ limit even within the mean field formalism, since requiring that $\alpha \ell / \beta$ is small is inconsistent with traditional estimates for the turbulent $\alpha$-effect ($\alpha \sim u$) and the turbulent $\beta$-effect ($\beta \sim \ell u$). In this case mean field theory predicts by equation (\ref{eq:grate}) that the fastest growing ``mean field'' has the same scale as the fluctuations --- namely a small-scale dynamo. It should be noted however that the growth rate of this small-scale dynamo predicted by mean field theory is not the correct one since it relies on lack of reflectional symmetry whereas
the actual growth rate does not.

All the considerations above address the kinematic evolution of magnetic fields, which formally is the relevant regime for mean field electrodynamics. However, there have been attempts to extend the mean field approach to the nonlinear regime. We conclude this paper by considering a particular case in which many of the issues we have discussed here are pertinent. It is possible to consider a case in which the velocity, rather than being prescribed, is the self-consistent solution of a saturated small-scale dynamo. Here both the velocity and magnetic fluctuations are stationary, and it is therefore possible, at least in principle, to measure $\alpha$ and $\beta$. In practice this process is tricky since any finite amplitude perturbation, such as the introduction of a mean field, will induce a corresponding change in the velocity. Putting aside the not inconsiderable technical difficulties involved in calculating $\alpha$ and $\beta$, it behoves us to ask whether these quantities convey useful information about magnetic field evolution.

To make these ideas concrete we consider a very specific case. First suppose that a numerical experiment is conducted on a periodic domain $L$ with $L \sim \ell$; we assume that small-scale dynamo action is observed, the magnetic field grows to a finite amplitude and saturates, and, by whatever means, $\alpha$ and $\beta$ are computed.\footnote{The problem is that we are interested in calculating $\alpha$ and $\beta$ as properties of the saturated nonlinear state. The measurement technique should not perturb the flow to a different state. Therefore of the techniques we described in \S3, probably only the Lagrangian approach can be guaranteed to achieve this result.} Now suppose that a new computational domain of size $NL$, where $N$ is a largish positive integer, is constructed by replication of the original domain. Clearly, in the absence of perturbation, the solution of the extended system will continue to evolve as $N^3$ replicas of the original system. If the system is subject to a long wavelength perturbation then it will relax to a new state, which, in general, is not periodic on scales smaller than $NL$; in other words, the system will transfer some of its energy to scales with wavenumbers smaller than $2 \pi /L$. The question is, will the coefficients $\alpha$ and $\beta$ capture any aspect of this relaxation process? The final state could be similar to the initial state (i.e.\ that obtained from replicating the smaller domain) or could be very different with, say, considerable energy at large scales. However, irrespective of the nature of the final state, the relaxation process by which it is achieved depends on the nonlinear interactions between all the scales larger than $L$ and is therefore not captured by averages over the scale $L$. Thus, in this case also, although $\alpha$ and $\beta$ can be measured, they do not convey useful information about magnetic field evolution.

\section*{Acknowledgments}

This research was supported by the National Science Foundation sponsored Center for Magnetic Self Organization (CMSO) at the University of Chicago, and by the Science and Technology Facilities Council. The paper was completed at the Kavli Institute for Theoretical Physics, supported by Grant No.\ NSF PHY05-51164.

\bsp

\label{lastpage}

\end{document}